\documentclass[a4paper,11pt]{article}

\usepackage[utf8]{inputenc}
\usepackage[margin=2cm]{geometry}
\usepackage{graphicx}
\usepackage{amsmath}
\usepackage{amssymb}
\usepackage{mathtools}
\usepackage{nicefrac}
\usepackage[braket]{qcircuit}
\usepackage{environ}
\newcommand{\myqctmp}[2][0.25]{\Qcircuit @C=#2em @R=#1em @!R}
\NewEnviron{myqcircuit}[1][0.25]{\vcenter{\myqctmp[#1]{0.5} {\BODY}}}

\usepackage{fancyvrb}
\RecustomVerbatimEnvironment{Verbatim}{Verbatim}{fontsize=\small}
\usepackage{pmboxdraw}
\usepackage{textcomp}
\usepackage{pifont}   
\usepackage{listings}
\usepackage[colorlinks,urlcolor=blue,linkcolor=black,citecolor=blue,breaklinks=true]{hyperref}
\usepackage[capitalise,nameinlink]{cleveref}

\title{QCLAB: A Matlab Toolbox for Quantum
Computing}
\author{
  Sophia Keip\thanks{FernUniversität in Hagen, Hagen, Germany. \texttt{sophia.keip@fernuni-hagen.de}} 
  \and 
  Daan Camps\thanks{Lawrence Berkeley National Laboratory (LBNL), Berkeley, USA. \texttt{dcamps@lbl.gov}, \texttt{rvanbeeumen@lbl.gov}} 
  \and 
  Roel Van Beeumen\footnotemark[2]
}
\date{March 2025}

\begin{document}

\maketitle

\begin{abstract}
We introduce QCLAB, an object-oriented MATLAB toolbox for constructing, representing, and simulating quantum circuits. Designed with an emphasis on numerical stability, efficiency, and performance, QCLAB provides a reliable platform for prototyping and testing quantum algorithms. For advanced performance needs, QCLAB\texttt{++} serves as a complementary C\texttt{++} package optimized for GPU-accelerated quantum circuit simulations. Together, QCLAB and QCLAB\texttt{++} form a comprehensive toolkit, balancing the simplicity of MATLAB scripting with the computational power of GPU acceleration. This paper serves as an introduction to the package and its features along with a hands-on tutorial that invites researchers to explore its capabilities right away.
\end{abstract}

\section{Introduction}
Quantum computing has rapidly advanced in recent years, propelled by remarkable progress in hardware development. As the technology evolves, the need for robust computational tools becomes ever more critical, empowering researchers to drive innovation in quantum algorithm research while hardware capabilities continue to mature.

This necessity led to the development of numerous quantum circuit simulators. Toolkits like IBM's Qiskit \cite{javadi2024quantum} with the Aer backend, Google Quantum AI's Cirq \cite{CirqDevelopers_2024} with the qsim simulator \cite{QuantumAI2020qsim} and Xanadu's PennyLane \cite{bergholm2018pennylane} offer advanced capabilities for quantum circuit design, each with their own strengths. Additionally, NVIDIA’s cuQuantum SDK \cite{bayraktar2023cuquantum} enables GPU-accelerated simulations, supported by all of the mentioned frameworks.

In this landscape, we present QCLAB\footnote{qclab: \url{https://github.com/QuantumComputingLab/qclab}} \cite{Camps2021QCLAB}, a MATLAB toolbox for efficiently designing, simulating and visualizing quantum circuits. What sets QCLAB apart is its MATLAB integration and its emphasis on numerical stability, efficiency, and performance. The focus on robust numerical techniques makes it especially useful for fields like numerical linear algebra and underlies its role as the foundational framework for a range of derived software packages and quantum compilers \cite{Camps2021F3C, camps2022fable,  Camps2022FABLE1}. Its intuitive syntax and integration with MATLAB offer flexibility and ease of use, making it highly accessible for both researchers and educators.

QCLAB uses the circuit model of quantum computation, which is widely adopted across quantum hardware platforms. By leveraging this model, QCLAB allows researchers to explore and prototype quantum algorithms efficiently on classical hardware, with seamless transition to real quantum hardware enabled through compatibility with OpenQASM.

QCLAB is complemented by its C\texttt{++} counterpart QCLAB\texttt{++}\footnote{qclab\texttt{++}: \url{https://github.com/QuantumComputingLab/qclabpp}} \cite{VanBeeumen2021QCLABpp, van2023qclab++}, which is designed for high-performance quantum circuit simulations on GPUs. Retaining the intuitive syntax of QCLAB, QCLAB\texttt{++} bridges the gap between MATLAB-based prototyping and GPU-accelerated large-scale simulations, providing a powerful toolkit for quantum algorithm research.

Beyond its computational capabilities, QCLAB includes intuitive tools for visualizing quantum circuits,  including the ability to generate executable LaTeX code. This simplifies the process of creating documentation for quantum research.

The paper is structured as follows: Section II demonstrates the intuitive quantum circuit construction in QCLAB, followed by Section III on simulation, combining practical steps with valuable insights into its implementation. Section IV discusses the visualization features, the openQASM compatibility, and the transition to QCLAB\texttt{++}. Section V continues with engaging examples, including quantum teleportation, state tomography, Grover's algorithm and quantum error correction. Finally, we conclude with a brief comparison to MATLAB's existing quantum computing package.

We assume a basic understanding of quantum computing principles, such as qubits, gates, and the circuit model. For a introduction we refer to \cite{nielsen2010quantum}.

\section{Constructing Quantum Circuits in QCLAB}
A quantum circuit represents quantum computations as a sequence of unitary operations (gates) applied to a register of qubits. It starts with an initial state, evolves through gates, and includes measurements that extract classical information. 
Quantum circuits are visually represented using musical score diagrams where qubits are shown as horizontal lines, while gates and measurements are symbols along these lines, with time proceeding from left to right. An example is illustrated in the circuit below.
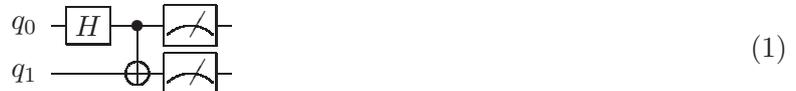
\begin{equation}
\begin{myqcircuit}
\lstick{q_{0}}	&	\gate{H}	&	\ctrl{1}	&	\meter	&	\qw	\\
\lstick{q_{1}}	&	\qw	&	\targ	&	\meter	&	\qw	\\
\end{myqcircuit}
\label{eq:circuit}
\end{equation}

With QCLAB, this concept can be translated into code efficiently. Let us construct the example circuit \eqref{eq:circuit} step by step. First, we create a \texttt{QCircuit} object by specifying the desired number of qubits. For a quantum circuit with 2 qubits, we write
\begin{center}
\begin{Verbatim}
            >> circuit = qclab.QCircuit(2);
\end{Verbatim}  
\end{center}
Once initialized, gates and measurements can be added to the circuit using the \texttt{push\_back} function. QCLAB provides a comprehensive implementation of commonly used quantum gates including controlled gates, leveraging their unitary matrix representations. Furthermore, its object-oriented architecture enables users to implement custom quantum gates, as demonstrated in \cite{Camps2021F3C}. In our small example circuit, the Hadamard gate and the Controlled-NOT (CNOT) gate are used. Each gate is specified with its target qubit or, in the case of controlled gates, both the control and target qubits as input parameters. For example the lines
\begin{center}
  \begin{Verbatim}
            >> circuit.push_back(qclab.qgates.Hadamard(0));
            >> circuit.push_back(qclab.qgates.CNOT(0,1));
  \end{Verbatim}
\end{center}
add a Hadamard gate to $q_0$ and a CNOT gate with control qubit $q_0$ and target qubit $q_1$. 

Measurements on $q_0$ and $q_1$ can be added as follows:
\begin{center}
  \begin{Verbatim}
            >> circuit.push_back(qclab.Measurement(0));
            >> circuit.push_back(qclab.Measurement(1));
  \end{Verbatim}
\end{center}
By default, measurements are performed in the \(Z\)-basis. However, \(X\)- and \(Y\)-basis measurements are preconfigured and can be specified as additional input arguments.
For custom bases, users can specify their own basis change operations. 

This straightforward approach allows for intuitive construction of quantum circuits, where gates and measurements are sequentially added to operate on the desired qubits.
Having illustrated how QCLAB enables the efficient construction of quantum circuits, we move on to the simulation of quantum circuits.

\section{Simulating Quantum Circuits in QCLAB}
Simulating quantum circuits is essential for testing and analyzing quantum algorithms before deploying them on actual hardware. QCLAB provides a full state vector simulation, accurately representing the entire quantum state and tracking amplitudes throughout the computation.

\subsection{Initialization} Every simulation of a quantum circuit requires an initial state. In QCLAB, this initial state can be specified either as a bitstring of zeros and ones, which is convenient for basis states, or as a vector, allowing for non-basis states. The simulation itself is performed using the \texttt{simulate} function. For example, to simulate the circuit \eqref{eq:circuit} with the initial state \(|00\rangle\), we write:
\begin{center}
 \begin{Verbatim}
            >> simulation = circuit.simulate('00');
 \end{Verbatim}   
\end{center}
Alternatively, if the vector representation is preferred, the simulation can be expressed as:
\begin{center}
 \begin{Verbatim}
            >> simulation = circuit.simulate([1;0;0;0]);
 \end{Verbatim}   
\end{center}

\subsection{Application of Gates} To apply a gate to a quantum state, QCLAB uses the unitary matrix associated with each gate object. Let \( U' \) represent the unitary matrix of a quantum gate acting on a subset of qubits. The corresponding unitary matrix \( U \), which operates on the entire register, is computed as:
\[
U = (I_l \otimes U' \otimes I_r),
\]
where $\otimes$ denotes the Kronecker product and \( I_l \) and \( I_r \) are identity matrices of appropriate dimensions, representing the unaffected qubits to the left and right of the gate’s target qubits. QCLAB computes a sparse representation of this extended unitary \( U \) for the entire register and applies it to the current state by multiplying it with the state vector. Optimized implementations of gate applications are implemented in QCLAB\texttt{++}~\cite{van2023qclab++}.

\subsection{Measurement} An essential aspect of circuit simulation is quantum measurement. Measurements in QCLAB are single-qubit operations.
The measurement process involves two steps: first, the probabilities of measuring the two possible outcomes $P(|0\rangle)$ and $P(|1\rangle)$ are computed using the amplitudes of the quantum state \( |\psi\rangle \), for example, 
\begin{align*}
    P(|0\rangle) = \sum_{i \in \{i \,|\, q_i = 0\}} |\psi_i|^2,
\end{align*}
where $\psi_i$ represents the amplitude of the basis state $|i\rangle$ and $q_i$ denotes the state of the measured qubit in $|i\rangle$.
Second, the state collapses to the subspace corresponding to the measurement result. 
For instance, if the outcome is \( |0\rangle \), the collapsed state is renormalized as:
\begin{align*}
|\psi_0\rangle = \frac{1}{\sqrt{P(|0\rangle)}} \sum_{i \in \{i \,|\, q_i = 0\}} \psi_i |i\rangle.
\end{align*}
Similarly, the same procedure applies for outcome \( |1\rangle \). In QCLAB, bitwise operations are used to efficiently determine the indices for constituting the collapsed state.

When measuring in a basis other than the computational basis, QCLAB applies the necessary basis changes prior to the standard measurement and reverts them afterward, ensuring accurate probabilities and post-measurement states for the specified basis. For example a measurement in the $X-$basis is implemented as
\begin{equation*}
\begin{myqcircuit}
&	\gate{H}	&	\meter	&	\gate{H^\dagger}&	 \qw \\
\end{myqcircuit}
\end{equation*}
When a measurement is performed and both outcomes have nonzero probabilities, the system is described by a probabilistic distribution over the possible post-measurement states. Each branch corresponds to one of the measurement outcomes, with its own collapsed state vector and associated probability. 

In our small example, there are two possible outcomes, each occurring with a probability of 0.5, resulting in two distinct post-measurement states. QCLAB makes this transparent by providing detailed results, probabilities, and collapsed states for each outcome:
\begin{center}
\begin{Verbatim}
            >> simulation.results
                ans =  2x1 cell
                    '00'
                    '11'
            >> simulation.probabilities
                ans =  2x1 cell
                    0.5000
                    0.5000
            >> simulation.states
                ans =  2x1 cell
                    [1;0;0;0]
                    [0;0;0;1]
\end{Verbatim}
\end{center}
For advanced use cases, such as quantum error correction (see \cref{sec:qec}) or iterative algorithms, QCLAB supports mid-circuit measurements and qubit resets \cite{decross2023qubit, ryan2022implementing}. After a mid-circuit measurement, the state evolution continues independently for each branch, ensuring an accurate representation of the quantum system across all possible measurement outcomes. 

If not all qubits are measured, QCLAB provides the capability to display the reduced state of the remaining unmeasured qubits, offering deeper insight into the system's partial state.

\section{Additional Features}
Other than the computational tasks, QCLAB enables the visualization of quantum circuits directly in the MATLAB command window
\begin{center}
  \begin{Verbatim}
            >> circuit.draw;
\end{Verbatim}  
\end{center}\vspace{-2ex}
\begin{lstlisting}
                    $  \mkern53.4mu┏━┓  \mkern41mu┏━┓ $
                    q0: $━┃\texttt{H}┃━\mkern2.8mu\bullet\mkern2.8mu━┃\texttt{M}┃━ $
                    $   \mkern53.4mu  ┗━┛\mkern17mu ┃ \mkern14.7mu┗━┛  $
                    $     \mkern98mu ┃\mkern14.1mu ┏━┓  $
                    q1: $━━━━━\oplus━┃\texttt{M}┃━ $
                    $    \mkern121.3mu   ┗━┛ $
\end{lstlisting}
which is very convenient to get an overview while constructing a quantum circuit.
Furthermore, QCLAB supports saving a circuit diagram to LaTeX source files, which was used to create all quantum circuits presented within this paper. This features can be accessed with the following command:
\begin{center}
  \begin{Verbatim}
            >> circuit.toTex();
\end{Verbatim}  
\end{center}
The effortless generation of executable LaTeX code makes it especially valuable for research documentation and presentations.
QCLAB also provides compatibility with openQASM, a low-level programming language used to describe quantum circuits, which is compatible with quantum hardware. This is accomplished using the command \texttt{toQASM}, which generates the following output for our circuit \eqref{eq:circuit}:
\begin{center}
  \begin{Verbatim}
            >> fID = 1; %file id
            >> circuit.toQASM(fID);
            
                    qreg q[2];
                    h q[0];
                    cx q[0], q[1];
                    measure q[0];
                    measure q[1];
\end{Verbatim}  
\end{center}
This allows users to test their quantum circuits on real quantum computers and bridges the gap between theory and implementation. 
Alongside the applications we present in the next section, QCLAB offers a variety of other examples that help users getting familiar with both quantum computing concepts and the toolbox itself. Additionally, documentation is available to make the learning process as accessible as possible.

We conclude this section by demonstrating how QCLAB and QCLAB\texttt{++} implementations share a similar structure, making it straightforward to convert between them. As already presented, creating a one-qubit circuit and adding a Hadamard gate to it in QCLAB is implemented by:

\begin{center}
  \begin{Verbatim}
            >> circuit = qclab.QCircuit(1);
            >> circuit.push_back(qclab.qgates.Hadamard(0));
  \end{Verbatim}
\end{center}
To accomplish the same in QCLAB\texttt{++}, the following code is required:
\begin{center}
  \begin{Verbatim}
            circuit = qclab::QCircuit<T>(1);
            circuit.push_back(
              std::make_unique<qclab::qgates::Hadamard<T>>(0)
            );
  \end{Verbatim}
\end{center}
The consistent programming interface enables a seamless workflow, allowing ideas to be prototyped in MATLAB and later scaled for GPU simulations using QCLAB\texttt{++}.

\section{Examples}
In this section, we showcase four standard quantum computing examples designed to help users get familiar with QCLAB. These examples are available in the package encouraging hands-on exploration and practical learning.

\subsection{Quantum Teleportation}
In this example, we demonstrate the simplicity of circuit construction in QCLAB, along with its ability to seamlessly integrate mid-circuit measurements and support arbitrary initial states.

Quantum teleportation transfers a state \(\ket{v}\) from one qubit (sender) to another (receiver) using entanglement and classical communication. The circuit diagram is shown below:
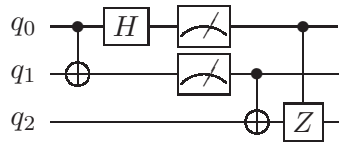
\begin{equation*}
\begin{myqcircuit}
\lstick{q_{0}}	&	\ctrl{1}	&	\gate{H}	&	\qw	&	\meter	&	\qw	&	\ctrl{2}	&	\qw	\\
\lstick{q_{1}}	&	\targ	&	\qw	&	\qw	&	\meter	&	\ctrl{1}	&	\qw	&	\qw	\\
\lstick{q_{2}}	&	\qw	&	\qw	&	\qw	&	\qw	&	\targ	&	\gate{Z}	&	\qw	\\
\end{myqcircuit}
\end{equation*}
Qubit \(q_0\) holds the state to be teleported, while \(q_1\) and \(q_2\) will be initialized as Bell pair providing the necessary entanglement. The qubits $q_0$ and $q_1$ are assumed to be in the possession of the sender, while $q_2$ is with the receiver. The sender performs a Bell measurement on \(q_0\) and \(q_1\) and transmits the results as classical bits. The receiver applies corrective operations (\(X\) and/or \(Z\), implemented as controlled gates), completing the teleportation of \(q_0\)'s state to \(q_2\). For more details, see \cite[Section 1.3.7]{nielsen2010quantum}.

Let us construct the quantum teleportation circuit, that we will call \texttt{qtc}, in QCLAB by first initializing the three-qubit circuit and afterward adding all gates and measurements:
\begin{center}
 \begin{Verbatim}
            >> qtc = qclab.QCircuit(3);
            >> qtc.push_back(qclab.qgates.CNOT(0,1));
            >> qtc.push_back(qclab.qgates.Hadamard(0));
            >> qtc.push_back(qclab.Measurement(0));   
            >> qtc.push_back(qclab.Measurement(1));      
            >> qtc.push_back(qclab.qgates.CNOT(1,2)); 
            >> qtc.push_back(qclab.qgates.CZ(0,2));   
\end{Verbatim}   
\end{center}

The simulation starts with an initial state, which is the tensor product of $\ket{v}= (\frac{1}{\sqrt{2}},i\frac{1}{\sqrt{2}})$, the state we chose to be teleported, and the Bell state:
\begin{center}
   \begin{Verbatim}
            >> v = [1/sqrt(2);1i/sqrt(2)];
            >> bell = [1/sqrt(2); 0; 0; 1/sqrt(2)];
            >> initial_state = kron(v, bell);
\end{Verbatim} 
\end{center}
QCLAB enables straightforward simulation of a quantum circuit on a specified initial state using the \texttt{simulate} function:
\begin{center}
\begin{Verbatim}
    >> simulation = qtc.simulate(initial_state);
\end{Verbatim}
\end{center}
To examine the outcomes of mid-circuit measurements, QCLAB provides easy access to the measurement results and their associated probabilities:

\begin{center}
\begin{Verbatim}
            >> simulation.results
            >> ans =
                4×1 cell array
                    {'00'}
                    {'01'}
                    {'10'}
                    {'11'}
            >> simulation.probabilities
            >> ans =
                    0.2500
                    0.2500
                    0.2500
                    0.2500
\end{Verbatim}
\end{center}

However, in this example the most interesting aspect is the quantum state at the end of the circuit. By using
\begin{center}
\begin{Verbatim}
            >> simulation.states
                  4×1 cell array
                    {8×1 double}
                    {8×1 double}
                    {8×1 double}
                    {8×1 double}
\end{Verbatim}
\end{center}
QCLAB returns the final state vector for each unique measurement result observed during the circuit. Here, the results are four distinct outcomes, so \texttt{simulation.states} contains four corresponding state vectors. The final state vector for the measurement outcome \texttt{'00'} reads
\begin{center}
\begin{Verbatim}
            >> simulation.states(1)
                ans =
                    0.7071 + 0.0000i
                    0.0000 + 0.7071i
                    0.0000 + 0.0000i
                    0.0000 + 0.0000i
                    0.0000 + 0.0000i
                    0.0000 + 0.0000i
                    0.0000 + 0.0000i
                    0.0000 + 0.0000i
\end{Verbatim}
\end{center}

To verify that the state \(\ket{v}\) was successfully teleported, QCLAB's built-in \texttt{reducedStatevector} function can be used to extract the state of qubit \(q_2\). The first argument is the final state vector, in the following we use \texttt{simulation.states(1)}, the second argument indicates which qubits are in a known state, here, that are the measured qubits $q_0$ and $q_1$, and the last argument indicates in which state the known qubits are, here \texttt{simulation.results(1) = '00'}.
\begin{center}
\begin{Verbatim}
            >> reducedStatevector(simulation.states(1),...
               [0,1], simulation.results(1))
                ans =
                    0.7071 + 0.0000i
                    0.0000 + 0.7071i
\end{Verbatim}
\end{center}
We see that the state reduced to the second qubit matches $|v\rangle$ ($\nicefrac{1}{\sqrt{2}}\approx 0.7071)$.
For end circuit measurements, where not the whole register is measured, the reduced states for the non measured qubits are included in the simulation and can be obtained by
\begin{center}
\begin{Verbatim}
            >> simulation.reducedStates
\end{Verbatim}
\end{center}
In this example, this is not applicable since we only have mid-circuit measurements.

\subsection{Quantum Tomography}
This example highlights the simplicity of performing measurements in different bases using QCLAB, while also demonstrating its capability to simulate repeated experiments - a fundamental aspect of quantum computing workflows.

Quantum tomography estimates the density matrix of an unknown quantum state by performing multiple measurements in various bases. Using QCLAB, we demonstrate the reconstruction of a one-qubit state \(\ket{v}\), given as state vector. The density matrix $\rho_v = \ket{v}\bra{v}$ of the state is estimated as
\begin{align}\label{equ:tomography}
  \rho_v^{\text{est}} = \frac{1}{2} \big(S_0 I + S_1 X + S_2 Y + S_3 Z \big),
\end{align}
where \(S_i\) are coefficients estimated from measurements in the $X$, $Y$, and $Z$-bases, \(I\) is the identity matrix and \( X, Y, Z\) are the Pauli matrices \cite{altepeter20044, nielsen2010quantum}.

To measure in the $X$-basis, we create a circuit and add a measurement operation:
\begin{center}
\begin{Verbatim}
            >> meas_x = qclab.QCircuit(1); 
            >> meas_x.push_back(qclab.Measurement(0, 'x'));
\end{Verbatim} 
\end{center}
Measurements in the $Y$- and $Z$-\,bases are set up analogously by specifying the respective basis as input to the \texttt{Measurement} object.

To estimate the density matrix, we measure the unknown state \(\ket{v}\) multiple times in three different bases. The accuracy of our estimate depends on the number of \texttt{shots}, which specifies how many times we measure. QCLAB simplifies this repetitive process with the \texttt{counts} function, where the number of shots is provided as input. This function returns the simulated frequency of each possible measurement outcome. 

For our single-qubit circuits, \texttt{counts} produces a two-element vector: the first entry represents the frequency of measuring 0, while the second indicates the rate of measuring 1. Below is the example for measurements in the $X$-basis simulated using $\ket{v} = (\frac{1}{\sqrt{2}},i\frac{1}{\sqrt{2}})$ as unknown state:
\begin{center}
\begin{Verbatim}
            >> res_x = meas_x.simulate(v);
            >> shots = 1000;
            >> rng(1) %setting a seed
            >> counts_x = res_x.counts(shots);
            >> counts_x =
                 471
                 529
\end{Verbatim}
\end{center}

The same process is repeated for measurements in the $Y$- and $Z$-bases. The estimated probabilities for measuring \( |0\rangle \), denoted as \( P^{\text{est}}_b(0) \), and \( |1\rangle \), denoted as \( P^{\text{est}}_b(1) \), in each basis \( b \), are calculated by dividing the count vector by the total number of shots. So for our example we get \( P^{\text{est}}_x(0) = \frac{471}{1000}=0.471 \) and \( P^{\text{est}}_x(1) = \frac{529}{1000}=0.529\) as estimates for the true probabilities \( P_x(0) = 0.5\) and \( P_x(1) = 0.5\).
 These probabilities are then used to compute the coefficients for the density matrix reconstruction:
\begin{align*}
S_0 &= P^{\text{est}}_z(0) + P^{\text{est}}_z(1)\\
S_1 &= P^{\text{est}}_x(0) - P^{\text{est}}_x(1), \\
S_2 &= P^{\text{est}}_y(0) - P^{\text{est}}_y(1), \\
S_3 &= P^{\text{est}}_z(0) - P^{\text{est}}_z(1). 
\end{align*}
This leads to $S_0=1$, $S_1=-0.058$, $S_2 = 1$ and $S_3 = -0.012$.
The coefficients are then substituted into \cref{equ:tomography} to construct the estimated density matrix \(\rho_v^{\text{est}}\) leading to
\begin{align*}
    \rho_v^{\text{est}} = \begin{bmatrix}
        0.494 & 0.029 -0.5i\\
        0.029 + 0.5i & 0.506
    \end{bmatrix}.
\end{align*}
In comparison, the actual density matrix $\rho_v$ is
\begin{align*}
    \rho_v = \begin{bmatrix}
        0.5 & -0.5i\\
        0.5i & 0.5
    \end{bmatrix}.
\end{align*}
The trace distance between $\rho_v$ and $\rho_v^{\text{est}}$ is $0.006$.

\subsection{Grover's Algorithm}
In this example, we demonstrate how QCLAB enables the modular construction of quantum circuits by combining smaller subcircuits into a larger, more complex circuit. 

Grover's algorithm provides a quadratic speedup for unstructured search problems, allowing a solution to be found in \(O(\sqrt{N})\), where \(N\) is the number of possible solutions. It achieves this by iteratively amplifying the probability of the desired solution using a combination of quantum superposition and interference. For more details on the algorithm, see \cite[Section 6.1]{nielsen2010quantum}.

Here, we use QCLAB to search for the state \(|11\rangle\) among four possibilities (\(|00\rangle\), \(|01\rangle\), \(|10\rangle\), and \(|11\rangle\)). Grover's algorithm consists of two key components: the oracle circuit and the diffuser circuit. The algorithm begins by initializing an equal superposition through the application of Hadamard gates. These are followed by the sequential application of the oracle and diffuser circuits, concluding with measurements, as illustrated in the following circuit diagram.
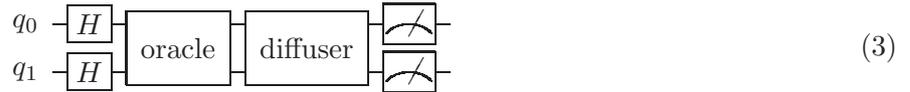
\begin{equation}
\begin{myqcircuit}
\lstick{q_{0}}	&	\gate{H}	&	\multigate{1}{\mathrm{oracle}}	&	\multigate{1}{\mathrm{diffuser}}	&	\meter	&	\qw	\\
\lstick{q_{1}}	&	\gate{H}	&	\ghost{\mathrm{oracle}}	&	\ghost{\mathrm{diffuser}}	&	\meter	&	\qw	\\
\end{myqcircuit}
\label{eq:grover}
\end{equation}
QCLAB allows for the independent construction of the oracle and diffuser circuits, which can then be combined into the complete Grover circuit. 

The oracle is responsible for flipping the phase of the marked state \(|11\rangle\). This can be easily implemented using a controlled-\(Z\) gate:
\begin{center}
\begin{Verbatim}
            >> oracle = qclab.QCircuit(2);
            >> oracle.push_back(qclab.qgates.CZ(0,1));
\end{Verbatim}
\end{center}
This results in the following oracle circuit:
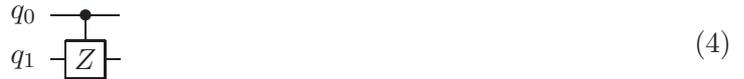
\begin{equation}\label{eq:oracle}
\begin{myqcircuit}
\lstick{q_{0}}	&	\ctrl{1}	&	\qw	\\
\lstick{q_{1}}	&	\gate{Z}	&	\qw	\\
\end{myqcircuit}
\end{equation}
The second component is the diffuser, which amplifies the amplitude of the marked state by reflecting the amplitudes about their average. 
The diffuser can be constructed as follows:
\begin{center}
\begin{Verbatim}
            >> diffuser = qclab.QCircuit(2);
            >> diffuser.push_back(qclab.qgates.Hadamard(0));
            >> diffuser.push_back(qclab.qgates.Hadamard(1));
            >> diffuser.push_back(qclab.qgates.PauliZ(0));
            >> diffuser.push_back(qclab.qgates.PauliZ(1));
            >> diffuser.push_back(qclab.qgates.CZ(0,1));
            >> diffuser.push_back(qclab.qgates.Hadamard(0));
            >> diffuser.push_back(qclab.qgates.Hadamard(1));
\end{Verbatim}
\end{center}
The corresponding circuit diagram is:
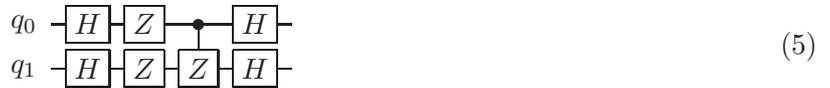
\begin{equation}\label{eq:diffuser}
\begin{myqcircuit}
\lstick{q_{0}}	&	\gate{H}	&	\gate{Z}	&	\ctrl{1}	&	\gate{H}	&	\qw	\\
\lstick{q_{1}}	&	\gate{H}	&	\gate{Z}	&	\gate{Z}	&	\gate{H}	&	\qw	\\
\end{myqcircuit}
\end{equation}
In order to draw the oracle and the diffuser circuits \eqref{eq:oracle} and \eqref{eq:diffuser} as blocks, like in \eqref{eq:grover}, we use
\begin{center}
\begin{Verbatim}
            >> oracle.asBlock;
            >> diffuser.asBlock;
\end{Verbatim}
\end{center}
which can be reversed by using \texttt{unBlock}.
With the two subcircuits prepared, we can now assemble the complete Grover circuit, called \texttt{gc}:
\begin{center}
\begin{Verbatim}
            >> gc = qclab.QCircuit(2);
            >> gc.push_back(qclab.qgates.Hadamard(0));
            >> gc.push_back(qclab.qgates.Hadamard(1));
            >> gc.push_back(oracle);
            >> gc.push_back(diffuser);
            >> gc.push_back(qclab.Measurement(0));
            >> gc.push_back(qclab.Measurement(1));
\end{Verbatim}
\end{center}
Next, we simulate the Grover circuit, beginning with the initial state \(|00\rangle\) and observe the results and probabilities:
\begin{center}
\begin{Verbatim}
            >> simulation = gc.simulate('00');
            >> simulation.results
            >> ans =
                1×1 cell array
                {'11'}
            >> simulation.probabilities
            >> ans =
                1.0000
\end{Verbatim}
\end{center}
The simulation yields \texttt{'11'} with a probability of 1, confirming the successful execution of the algorithm.

\subsection{Quantum Error Correction}\label{sec:qec}
In this example, we demonstrate another use case that requires
mid-circuit measurements. Additionally, we showcase the implementation of multi-controlled gates in QCLAB.

Quantum error correction (QEC) protects quantum information from errors during computation by encoding the logical information across multiple physical qubits~\cite[Section 10.1.1]{nielsen2010quantum}. In this example, we demonstrate the detection and correction of a single bit-flip error using a distance-3 repetition code. Our example requires a 5-qubit circuit:
\begin{center}
\begin{Verbatim}
            >> qec = qclab.QCircuit(5);
\end{Verbatim}
\end{center}
Let us start with the single-qubit state we want to protect:
\begin{align*}
 |v\rangle = \alpha |0\rangle + \beta |1\rangle.   
\end{align*}
The first step is to encode this state in three qubits, called the physical qubits, as 
\begin{align*}
 \ket{v}_L = \alpha |000\rangle + \beta |111\rangle.
\end{align*}
Using this encoding, the three physical qubits are combined to act as a logical qubit state $\ket{v}_L$ that is protected against a single bit-flip error. A bit-flip error can be detected by majority voting.
The state $\ket{v}_L$ can be generated by the following circuit:

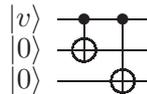
\begin{equation*}
\begin{myqcircuit}
\lstick{|v\rangle}	&	\ctrl{1}	&	\ctrl{2} &	\qw\\
\lstick{|0\rangle}	&	\targ	&	\qw	&	\qw\\
\lstick{|0\rangle}	&	\qw	&	\targ	&	\qw
\end{myqcircuit}
\end{equation*}
The physical qubits correspond to the first 3 qubits, $q_0$, $q_1$ and $q_2$, of the \texttt{qec} circuit, so we need to add the following sequence of CNOT gates to the circuit:
\begin{center}
\begin{Verbatim}
            >> qec.push_back(qclab.qgates.CNOT(0,1));
            >> qec.push_back(qclab.qgates.CNOT(0,2));
\end{Verbatim}
\end{center}

Next, we introduce a bit-flip error on the first physical qubit $q_0$ by adding a Pauli-$X$ gate:
\begin{center}
\begin{Verbatim}
            >> qec.push_back(qclab.qgates.PauliX(0));
\end{Verbatim}
\end{center}
This leaves the physical qubits in the state 
\begin{align*}
\alpha |100\rangle + \beta |011\rangle   
\end{align*}
 which lies outside the logical subspace. A measurement of the qubits $q_0$, $q_1$ and $q_2$ would reveal the bit-flip error, but would simultaneously destroy the encoded logical qubit state.
Instead, QEC utilizes ancilla qubits to
extract the so-called error syndrome, which detects the error without destroying the state of the logical qubit. In our example, we use $q_3$ and $q_4$ as ancillas to detect the bit-flip error. To do so, we apply a series of CNOT gates, which entangle the logical qubits with the ancilla qubits, extracting information about the error into the ancilla states:
\begin{center}
\begin{Verbatim}
            >> qec.push_back(qclab.qgates.CNOT(0,3));
            >> qec.push_back(qclab.qgates.CNOT(1,3));
            >> qec.push_back(qclab.qgates.CNOT(0,4));
            >> qec.push_back(qclab.qgates.CNOT(2,4));
\end{Verbatim}
\end{center}
Next, the ancilla qubits $q_3$ and $q_4$ are measured to determine if a bit-flip error occurred and, if so, on which qubit. These measurements provide the error syndrome, which guides the correction process without collapsing the logical qubit's encoded state:
\begin{center}
\begin{Verbatim}
            >> qec.push_back(qclab.Measurement(3));
            >> qec.push_back(qclab.Measurement(4));
\end{Verbatim}
\end{center}
Using the error syndrome obtained from the ancilla qubits, we add three multi-controlled $X$-gates\footnote{In practice, the correction can be implemented more efficiently using Clifford gates and classical control, or even entirely in software by tracking the Pauli frame. This removes the need for coherent multi-controlled gates.}, with the ancilla qubits serving as controls. These gates target the affected qubit and correct the detected bit-flip error, effectively restoring the protected quantum state:
\begin{center}
\begin{Verbatim}
            >> qec.push_back(qclab.qgates.MCX([3,4],2,[0,1]))
            >> qec.push_back(qclab.qgates.MCX([3,4],1,[1,0]))
            >> qec.push_back(qclab.qgates.MCX([3,4],0,[1,1]))
\end{Verbatim}
\end{center}
%
%
%
The complete circuit is displayed below:
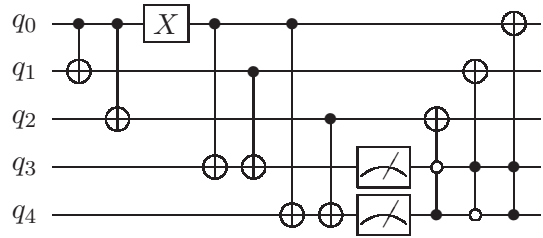
\begin{equation*}
\begin{myqcircuit}
\lstick{q_{0}}	&	\ctrl{1}	&	\ctrl{2}	&	\gate{X}	&	\ctrl{3}	&	\qw	&	\ctrl{4}	&	\qw	&	\qw	&	\qw	&	\qw	&	\targ	&	\qw \\
\lstick{q_{1}}	&	\targ	&	\qw	&	\qw	&	\qw	&	\ctrl{2}	&	\qw	&	\qw	&	\qw	&	\qw	&	\targ	&	\qw	&	\qw \\
\lstick{q_{2}}	&	\qw	&	\targ	&	\qw	&	\qw	&	\qw	&	\qw	&	\ctrl{2}	&	\qw	&	\targ	&	\qw	&	\qw	&	\qw		\\
\lstick{q_{3}}	&	\qw	&	\qw	&	\qw	&	\targ	&	\targ	&	\qw	&	\qw	&	\meter	&	\ctrlo{-1}	&	\ctrl{-2}	&	\ctrl{-3}	&	\qw	\\
\lstick{q_{4}}	&	\qw	&	\qw	&	\qw	&	\qw	&	\qw	&	\targ	&	\targ	&	\meter	&	\ctrl{-1}	&	\ctrlo{-1}	&	\ctrl{-1}	&	\qw	\\
\end{myqcircuit}
\end{equation*}
Let us check if this circuit protects the state $|v\rangle = (\frac{1}{\sqrt{2}},i\frac{1}{\sqrt{2}})$ by simulating it.
\begin{center}
\begin{Verbatim}
            >> % creating initial state
            >> v = [1/sqrt(2); 1i/sqrt(2)];
            >> initial_state = kron(v,eye(16,1));
            
            >> simulation = qec.simulate(initial_state);
            >> simulation.results
                ans =
                 1×1 cell array
                {'11'}
            >> simulation.probabilities
                ans =
                 1
\end{Verbatim}
\end{center}
The measurement result \texttt{'11'} indicates that the third correcting multi-controlled $X$-gates was executed. This gate reversed the bit flip introduced on the first qubit, precisely achieving the intended error correction and restoring the physical qubits to their correct state.

\section{Conclusion}

In this paper, we introduced QCLAB, an object-oriented MATLAB toolbox for constructing, simulating, and visualizing quantum circuits. Through four diverse examples, we demonstrated the versatility and practicality of QCLAB, encouraging readers to explore its features firsthand.

QCLAB enables straightforward prototyping of quantum algorithms for experienced researchers while providing an easy entry point into quantum computing for newcomers, making it suitable for a wide audience.

By mirroring the syntax of QCLAB\texttt{++}, QCLAB allows for an efficient transition from prototyping to high-performance and GPU-accelerated simulations, bridging the gap to large-scale computations.

 QCLAB sets itself apart from MATLAB’s own quantum computing package by offering an open-source and object-oriented architecture, enabling users to implement own functionalities such as custom quantum gates. It further supports mid-circuit and partial measurements, measurements in arbitrary bases, offers LaTeX export for circuit diagrams, and seamlessly translates to QCLAB\texttt{++} for GPU-accelerated simulations. These capabilities go beyond MATLAB’s current solution, making QCLAB an advanced tool for both prototyping and quantum computing research.

\bibliographystyle{abbrvurl} 
\bibliography{ref}

\begin{thebibliography}{10}

\bibitem{altepeter20044}
J.~B. Altepeter, D.~F. James, and P.~G. Kwiat.
\newblock 4 qubit quantum state tomography.
\newblock In {\em Quantum state estimation}, pages 113--145. Springer, 2004.
\newblock \href {https://doi.org/10.1007/978-3-540-44481-7_4} {\path{doi:10.1007/978-3-540-44481-7_4}}.

\bibitem{bayraktar2023cuquantum}
H.~Bayraktar, A.~Charara, D.~Clark, S.~Cohen, T.~Costa, Y.-L.~L. Fang, Y.~Gao, J.~Guan, J.~Gunnels, A.~Haidar, et~al.
\newblock cuquantum sdk: A high-performance library for accelerating quantum science.
\newblock In {\em 2023 IEEE International Conference on Quantum Computing and Engineering (QCE)}, volume~1, pages 1050--1061. IEEE, 2023.
\newblock \href {https://doi.org/10.1109/QCE57702.2023.00119} {\path{doi:10.1109/QCE57702.2023.00119}}.

\bibitem{bergholm2018pennylane}
V.~Bergholm, J.~Izaac, M.~Schuld, C.~Gogolin, S.~Ahmed, V.~Ajith, M.~S. Alam, G.~Alonso-Linaje, B.~AkashNarayanan, A.~Asadi, et~al.
\newblock Pennylane: Automatic differentiation of hybrid quantum-classical computations.
\newblock {\em arXiv preprint}, 2018.
\newblock \href {https://doi.org/10.48550/arXiv.1811.04968} {\path{doi:10.48550/arXiv.1811.04968}}.

\bibitem{Camps2021QCLAB}
D.~Camps, S.~Keip, and R.~{Van Beeumen}.
\newblock {QCLAB v1.0.0}, mar 2025.
\newblock Available on GitHub.
\newblock URL: \url{https://github.com/QuantumComputingLab/qclab}, \href {https://doi.org/10.5281/zenodo.14968124} {\path{doi:10.5281/zenodo.14968124}}.

\bibitem{Camps2021F3C}
D.~Camps and R.~{Van Beeumen}.
\newblock {F3C}, 2021.
\newblock Available on GitHub.
\newblock URL: \url{https://github.com/QuantumComputingLab/f3c}, \href {https://doi.org/10.5281/zenodo.5160760} {\path{doi:10.5281/zenodo.5160760}}.

\bibitem{camps2022fable}
D.~Camps and R.~Van~Beeumen.
\newblock {FABLE: Fast approximate quantum circuits for block-encodings}.
\newblock In {\em 2022 IEEE International Conference on Quantum Computing and Engineering (QCE)}, pages 104--113. IEEE, 2022.
\newblock \href {https://doi.org/10.1109/QCE53715.2022.00029} {\path{doi:10.1109/QCE53715.2022.00029}}.

\bibitem{Camps2022FABLE1}
D.~Camps and R.~{Van Beeumen}.
\newblock {Fast Approximate BLock Encodings (FABLE) v0.1.1}, 2022.
\newblock Available on GitHub.
\newblock URL: \url{https://github.com/QuantumComputingLab/fable}.

\bibitem{CirqDevelopers_2024}
{Cirq Developers}.
\newblock {\em Cirq}.
\newblock Zenodo, May 2024.
\newblock URL: \url{https://zenodo.org/doi/10.5281/zenodo.4062499}, \href {https://doi.org/10.5281/ZENODO.4062499} {\path{doi:10.5281/ZENODO.4062499}}.

\bibitem{decross2023qubit}
M.~DeCross, E.~Chertkov, M.~Kohagen, and M.~Foss-Feig.
\newblock Qubit-reuse compilation with mid-circuit measurement and reset.
\newblock {\em Physical Review X}, 13(4):041057, 2023.
\newblock \href {https://doi.org/10.1103/PhysRevX.13.041057} {\path{doi:10.1103/PhysRevX.13.041057}}.

\bibitem{javadi2024quantum}
A.~Javadi-Abhari, M.~Treinish, K.~Krsulich, C.~J. Wood, J.~Lishman, J.~Gacon, S.~Martiel, P.~D. Nation, L.~S. Bishop, A.~W. Cross, et~al.
\newblock Quantum computing with qiskit.
\newblock {\em arXiv preprint}, 2024.
\newblock \href {https://doi.org/10.48550/arXiv.2405.08810} {\path{doi:10.48550/arXiv.2405.08810}}.

\bibitem{nielsen2010quantum}
M.~A. Nielsen and I.~L. Chuang.
\newblock {\em Quantum computation and quantum information}.
\newblock Cambridge university press, 2010.
\newblock \href {https://doi.org/10.1017/CBO9780511976667} {\path{doi:10.1017/CBO9780511976667}}.

\bibitem{QuantumAI2020qsim}
{Quantum AI team and collaborators}.
\newblock qsim, sep 2020.
\newblock Available on Zenodo.
\newblock \href {https://doi.org/10.5281/zenodo.4023103} {\path{doi:10.5281/zenodo.4023103}}.

\bibitem{ryan2022implementing}
C.~Ryan-Anderson, N.~Brown, M.~Allman, B.~Arkin, G.~Asa-Attuah, C.~Baldwin, J.~Berg, J.~Bohnet, S.~Braxton, N.~Burdick, et~al.
\newblock Implementing fault-tolerant entangling gates on the five-qubit code and the color code.
\newblock {\em arXiv preprint}, 2022.
\newblock \href {https://doi.org/10.48550/arXiv.2208.01863} {\path{doi:10.48550/arXiv.2208.01863}}.

\bibitem{VanBeeumen2021QCLABpp}
R.~{Van Beeumen} and D.~Camps.
\newblock {QuantumComputingLab/qclabpp: QCLAB++ v0.1.2}, aug 2021.
\newblock Available on GitHub.
\newblock URL: \url{https://github.com/QuantumComputingLab/qclabpp}, \href {https://doi.org/10.5281/zenodo.5160682} {\path{doi:10.5281/zenodo.5160682}}.

\bibitem{van2023qclab++}
R.~Van~Beeumen, D.~Camps, and N.~Mehta.
\newblock {QCLAB++: Simulating Quantum Circuits on GPUs}.
\newblock {\em arXiv preprint}, 2023.
\newblock \href {https://doi.org/10.48550/arXiv.2303.00123} {\path{doi:10.48550/arXiv.2303.00123}}.

\end{thebibliography}

\end{document}